\documentclass[prl,twocolumn,showpacs]{revtex4-1}

\usepackage{graphicx}
\usepackage{times}
\usepackage{amsmath,amssymb}

\newcommand{\Na}{Na$_2$IrO$_3$}
\newcommand{\Li}{Li$_2$IrO$_3$}
\newcommand{\CW}{$\Lambda_{\rm CW}$}

\begin{document}

\title{Finite-temperature phase diagram of the Heisenberg-Kitaev model}

\author{Johannes Reuther${}^1$}
\author{Ronny Thomale${}^{2,3}$}
\author{Simon Trebst${}^{3}$}

\affiliation{${}^1$Institut f\"ur Theorie der Kondensierten Materie,
  Karlsruhe Institute of Technology, 76128 Karlsruhe, Germany}
\affiliation{${}^2$Department of Physics, Princeton University, Princeton, NJ 08544, USA}
\affiliation{${}^3$Microsoft Research, Station Q, University of California, Santa Barbara, CA 93106}

\date{\today}

\begin{abstract}
We discuss the finite-temperature phase diagram of the Heisenberg-Kitaev model
on the hexagonal lattice, which has been suggested to describe the spin-orbital exchange 
of the effective spin-1/2 momenta in the Mott insulating Iridate \Na.
At zero-temperature this model exhibits magnetically ordered states well beyond the 
isotropic Heisenberg limit as well as an extended gapless spin liquid phase around 
the highly anisotropic Kitaev limit.
Using a pseudofermion functional renormalization group (RG) approach we extract
both the Curie-Weiss scale and the critical ordering scale (for the magnetically ordered 
states) from the RG flow of the magnetic susceptibility. 
The Curie-Weiss scale switches sign -- indicating a transition of the dominant exchange from antiferromagnetic
to ferromagnetic -- deep in the magnetically ordered regime. 
For the latter we find no significant frustration, i.e. a substantial suppression 
of the ordering scale with regard to the Curie-Weiss scale.
We discuss our results in light of recent experimental susceptibility measurements for \Na.
\end{abstract}

\pacs{71.20.Be, 75.25.Dk, 75.30.Et, 75.10.Jm}

\maketitle


In the realm of solid state physics, frustration refers to the phenomena that arise from the competition 
between interactions that cannot be simultaneously satisfied: typically a large degeneracy of ground states and 
a suppression of thermal ordering by fluctuations~\cite{Ramirez}. 
For many magnetic solids a peculiar form of frustration, so-called geometric frustration, 
can arise when interactions are incompatible with the underlying lattice symmetry~\cite{Moessner}. 
A prominent example of the latter are 
spin-1/2 Heisenberg antiferromagnets on non-bipartite lattice 
structures, for which there is no straight-forward generalization of the N\'eel state -- the common 
ground state for bipartite lattices -- but which can instead harbor more exotic ground states, 
including commonly elusive spin liquids~\cite{Balents}. 
Even for bipartite lattices one can encounter geometric frustration when considering so-called
orbital degrees of freedom, which occur in a large class of transition metal oxides that exhibit
Jahn-Teller ions~\cite{Khomskii}. For the latter crystal field splitting often results
in a single electron (or hole) occupying the doubly degenerate e$_{\rm g}$ level, for which the orbital
occupation is then cast in terms of a pseudospin-$1/2$. 
In contrast to ordinary spin degrees of freedom the exchange interactions between these orbital 
degrees of freedom -- arising from Jahn-Teller distortions and/or superexchange -- are highly 
anisotropic and even for simple bipartite lattices cannot be simultaneously satisfied, which has been
shown to result, {\em e.g.} in a non-trivial phase diagram of competing orbital orders on the cubic lattice \cite{Rynbach} 
or an orbital Coulomb phase on the diamond lattice \cite{ChernWu}.

In this manuscript, we consider a class of materials, certain Iridates, where strong spin-orbit coupling 
(SOC) results in effective degrees of freedom, which fall between the two opposing cases above. 
While Iridates have attracted much recent attention as candidate materials for topological 
insulators~\cite{TopologicalInsulators}, our study is motivated by a family of materials of the 
form A$_2$IrO$_3$, such as \Na, which has recently been shown to be a Mott insulator~\cite{singh-10prb064412}. 
In these Iridates the Ir$^{4+}$ ($5d^5$) ions form a
quasi two-dimensional hexagonal lattice of effective $j=1/2$ momenta. The latter arise from the
combined effect of crystal field splitting of the $d$-orbitals, resulting in a single hole (5 electrons)
occupying the lowered t$_{\rm 2g}$ orbitals, and spin-orbit coupling then giving rise to two Kramers
doublets, four electrons filling the (lower) $j=3/2$ quartet and a single electron in the $j=1/2$ doublet. 
The exchange interactions between these effective moments have been argued \cite{jackeli-09prl017205,chaloupka-10prl027204}
to reflect both the original spin exchange in terms of an isotropic Heisenberg coupling as well as
strongly anisotropic orbital interactions in terms of a Kitaev-type exchange
\begin{equation}
   H_{\rm HK}[\alpha] =  (1-\alpha) {\sum_{\langle i,j\rangle}} \vec{\sigma}_i\cdot \vec{\sigma}_j
                        - 2\alpha \sum_{{\gamma \rm-links}} {\sigma_i^{\gamma} \sigma_j^{\gamma}} \,,
   \label{Eq:KitaevHeisenbergHamiltonian}
\end{equation}
where the $\sigma_i$ denote the effective spin-1/2 moment of the Ir$^{4+}$ ions and $\gamma = x,y,z$ indicates
the three different links of the hexagonal lattice.
The two couplings are found \cite{chaloupka-10prl027204} to enter with opposite sign, i.e. the isotropic 
exchange is antiferromagnetic, while the anisotropic exchange is ferromagnetic.
Varying the relative coupling strength $0 \leq \alpha \leq 1$,
the model interpolates from the ordinary Heisenberg model with a N\'eel ground state for $\alpha=0$ 
to the Kitaev model for $\alpha=1$, which even for ferromagnetic interactions is highly frustrated and exhibits 
a gapless spin-liquid ground state \cite{Kitaev06ap2}.
One might thus wonder how the level of frustration varies between the spin and orbital dominated
limits of this model. This question is also fueled by  recent experiments~\cite{singh-10prb064412} 
on \Na\ that reported magnetic susceptibility measurements, which besides providing unambiguous
evidence of the effective spin-1/2 moments also reported a considerable suppression for the onset of 
magnetic correlations below $T_{\text{N}} \approx 15$~K in comparison with a Curie-Weiss
temperature of $\Theta_{\text{CW}} \approx -116$~K. In particular, one might wonder whether this
suppression of magnetic ordering might be interpreted as arising from a proximity to the highly exotic 
spin liquid phase of the Kitaev model, despite recent resonant x-ray magnetic scattering experiments~\cite{Hill} 
reporting indications of a conventionally ordered magnetic ground state.

\begin{figure*}[t]
    \includegraphics[width=\linewidth]{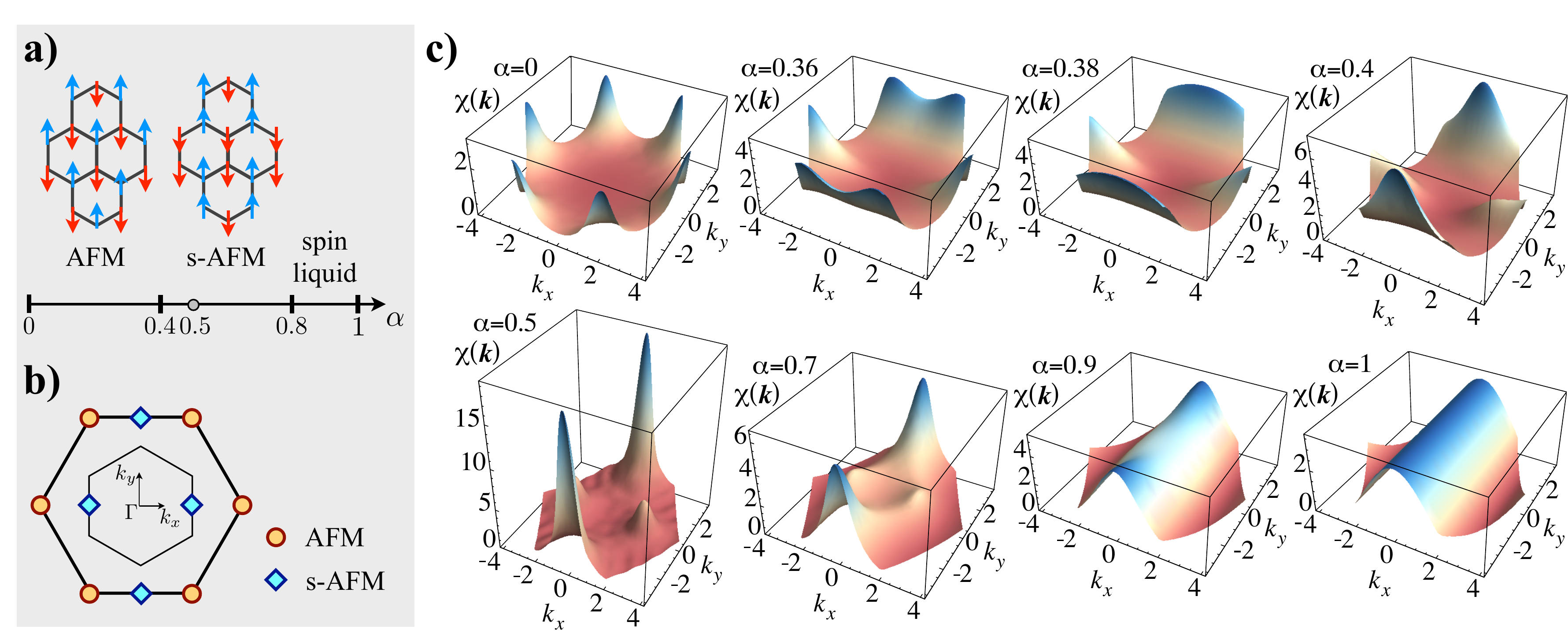}
   \caption{(color online) 
                   a) Zero-temperature phase diagram of the Heisenberg-Kitaev model \eqref{Eq:KitaevHeisenbergHamiltonian} exhibiting 
                        two magnetically ordered (AFM, s-AFM) and a spin liquid phase. 
                   b) Positions of the ordering peaks for the magnetically ordered phases in the extended Brillouin zone (BZ). 
                        The inner hexagon indicates the first BZ. 
                        Note that s-AFM order possesses two inequivalent peak positions in the extended BZ. 
                   c) Evolution of the $k$-space resolved static magnetic susceptibility upon variation of the coupling parameter $\alpha$. 
}
   \label{FRG-chi}
\end{figure*}

In this paper, we address the above questions by investigating the finite-temperature phase diagram 
of model~\eqref{Eq:KitaevHeisenbergHamiltonian}.
We use a recently developed pseudofermion functional renormalization group
(PF-FRG)~\cite{reuther-10prb144410,reuther-10prb024402,reuther-cm1103}
approach to compute the
magnetic susceptibility from the pseudofermion two-particle vertex
function evolving under an RG flow with a frequency cutoff $\Lambda$. 
Neglecting particle fluctuations induced by
thermal fluctuations, i.e. assuming that we are in the strong coupling
limit where the on-site repulsion exceeds the temperature, we argue
and numerically substantiate that the PF-FRG provides a suitable tool
to obtain both finite-temperature and ground-state properties of the 
model allowing for a direct comparison to thermodynamic experiments. 
In particular, we extract the high-temperature Curie-Weiss
behavior from the RG flow, the onset of magnetic ordering  (from the 
breakdown of the RG flow) and momentum-resolved magnetic susceptibility 
profiles, which also allow to identify the nature of the various ground 
states of model \eqref{Eq:KitaevHeisenbergHamiltonian}.

\paragraph{\it Numerical simulations.--}
The PF-FRG approach~\cite{reuther-10prb144410,reuther-10prb024402,reuther-cm1103}
starts by reformulating the spin Hamiltonian in terms
of a pseudo\-fermion representation of the spin-1/2 operators 
$S^{\mu} = 1/2 \sum_{\alpha\beta} f_{\alpha}^{\dagger} \sigma_{\alpha\beta}^{\mu} f_{\beta}$,
($\alpha,\beta = \uparrow,\downarrow$, $\mu = x,y,z$) with 
fermionic operators $f_{\uparrow}$ and $f_{\downarrow}$ and 
Pauli-matrices $\sigma^{\mu}$. 
Such a representation enables one to apply Wick's theorem leading to
standard Feynman many-body techniques.
In this pseudofermion language, quantum spin models become strong-coupling models 
with zero fermionic bandwidth and a finite interaction strength. The major conceptual
advancement of the FRG approach~\cite{wetterich93plb90,shankar94rmp129,honerkamp-01prb035109,reuther-10prb144410,reuther-10prb024402}
is that it allows to tackle this situation by providing a systematic scheme for the
(infinite) resummations needed in a direct perturbative attack. These FRG summations
are obtained in different interaction channels by generating equations for the evolution 
of all $m$-particle vertex functions under the flow of a sharp infrared 
frequency cutoff $\Lambda$. 
To reduce the infinite hierarchy of equations to a closed set, a common approach is to 
restrict oneself to one-loop diagrams. The PF-FRG extends this approach by also including certain 
two-loop contributions~\cite{katanin04prb115109} to retain a sufficient backfeeding of the self-energy
corrections to the two-particle vertex evolution. 
It is this two-particle vertex, which allows to compute the magnetic susceptibility -- our
main diagnostic tool to study 
model \eqref{Eq:KitaevHeisenbergHamiltonian}.

The FRG equations are simultaneously solved on the imaginary frequency axis and 
in real space. A numerical solution requires to i) discretize the frequency dependencies
and ii) to limit the spatial dependence to a finite cluster, thus keeping correlations
only up to some maximal length. In our calculations the latter typically extends over 7 lattice 
spacings corresponding to a correlation area (cluster size) of 112 lattice sites for the hexagonal 
lattice at hand. 
The onset of spontaneous long-range order is signaled by a sudden breakdown of the smooth 
RG flow, while the existence of a stable solution indicates the absence of long-range
order, see Refs.~\onlinecite{reuther-10prb144410,reuther-10prb024402} for further technical
details.


\paragraph{\it Zero-temperature states.--}

We start our discussion by first considering the zero-temperature phases 
of the Heisenberg-Kitaev model \eqref{Eq:KitaevHeisenbergHamiltonian}
and by recapitulating previous $T=0$ results~\cite{chaloupka-10prl027204}.
Interpolating the relative coupling strength $\alpha$
between the  Heisenberg limit ($\alpha=0$) and the Kitaev limit ($\alpha=1$), 
a sequence of three phases has
been observed~\cite{chaloupka-10prl027204}: The N\'eel ordered (AFM)
state of the Heisenberg limit is stable for $\alpha \lesssim 0.4$,
when it gives way to a `stripy' N\'eel ordered (s-AFM) state
illustrated in Fig.~\ref{FRG-chi}a) which covers the coupling regime $0.4
\lesssim \alpha \lesssim 0.8$.  In the extended parameter regime $0.8
\lesssim \alpha \leq 1$ the collective ground state is a gapless spin
liquid (SL) where the emerging gapless excitations are Majorana fermions 
forming two Dirac cones in momentum space~\cite{Kitaev06ap2}.

\begin{figure}[t]
    \includegraphics[width=\columnwidth]{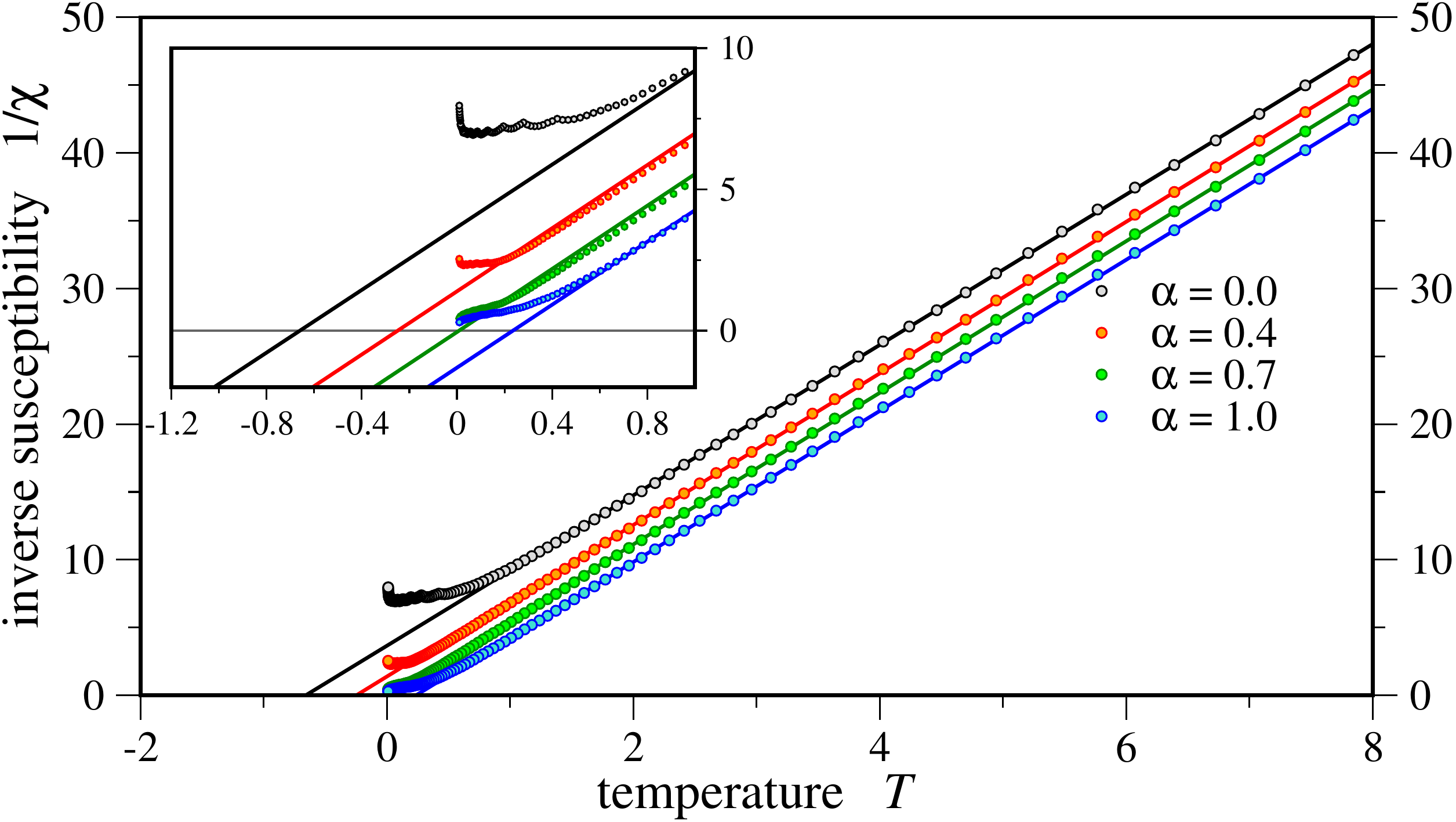}
   \caption{(color online)
              The inverse susceptibility $1/\chi$ as obtained from pseudofermion FRG calculations
              for various coupling parameters $\alpha$.
              The solid lines indicate fits to a Curie-Weiss law \eqref{Eq:CurieWeiss}.
             }
   \label{Fig:CW}
\end{figure}

We have calculated the characteristic magnetic susceptibility profiles for these 
states within our PF-FRG approach, as given for various values of $\alpha$ 
in the main panel of Fig.~\ref{FRG-chi}, where we plot the susceptibility just above 
the (finite) ordering scale $\Lambda_c$ below which the RG evolution becomes unstable. 
We adopt the extended Brillouin zone (BZ), illustrated in Fig.~\ref{FRG-chi}b),
appropriate for the two-site unit cell of the hexagonal lattice. 
For the N\'eel ordered phase we observe characteristic corner peaks in the susceptibility.
This magnetic AFM signature remains robust for the full extent of the phase
up to $\alpha \approx 0.38$ where we observe a relatively abrupt shift of the
susceptibility maxima. The latter is indicative of a first-order phase
transition, which is in tune with previous $T=0$ numerical studies
\cite{chaloupka-10prl027204,jiang-cm1101}.
Above $\alpha \approx 0.38$ we observe the onset of the second magnetically
ordered phase, the stripy AFM, for which the susceptibility 
signature comes in the form of {\em two} (not symmetry related) maxima, 
with a dominant peak along $k_x=0$ in the second BZ and a smaller peak along 
$k_y=0$ in the first BZ.
From an extrapolation of the finite-temperature crossover line that separates dominant 
AFM and dominant s-AFM fluctuation regimes, we can locate the zero temperature transition at
$\alpha\approx0.4$ in correspondence with previous studies \cite{chaloupka-10prl027204,jiang-cm1101}. 
Within the s-AFM phase, the point $\alpha=0.5$ stands out for which
the exact quantum ground-state has been shown \cite{chaloupka-10prl027204}
to be the classically ordered state (without any dressing). 
In our calculations, the absence of (quantum) fluctuations at this point is indicated 
by remarkably sharp response peaks, see Fig.~\ref{FRG-chi}c).
For $\alpha > 0.5$ we observe a pronounced decrease of the peaks. 
In contrast to the transition between the two magnetically ordered phases, 
the phase transition from the s-AFM phase to the spin liquid phase around the 
Kitaev limit is more subtle to detect in our calculations. 
Around the previously reported quantum critical point \cite{chaloupka-10prl027204,jiang-cm1101} 
at $\alpha \approx 0.8$ we observe a smooth evolution of the susceptibility profile 
into the one expected for the SL phase: a pure $\cos$-type susceptibility, reminiscent of the 
purely nearest-neighbor correlations in this phase \cite{Baskaran}. 


\paragraph{\it Finite-temperature physics.--}

\begin{figure}[t]
    \includegraphics[width=\columnwidth]{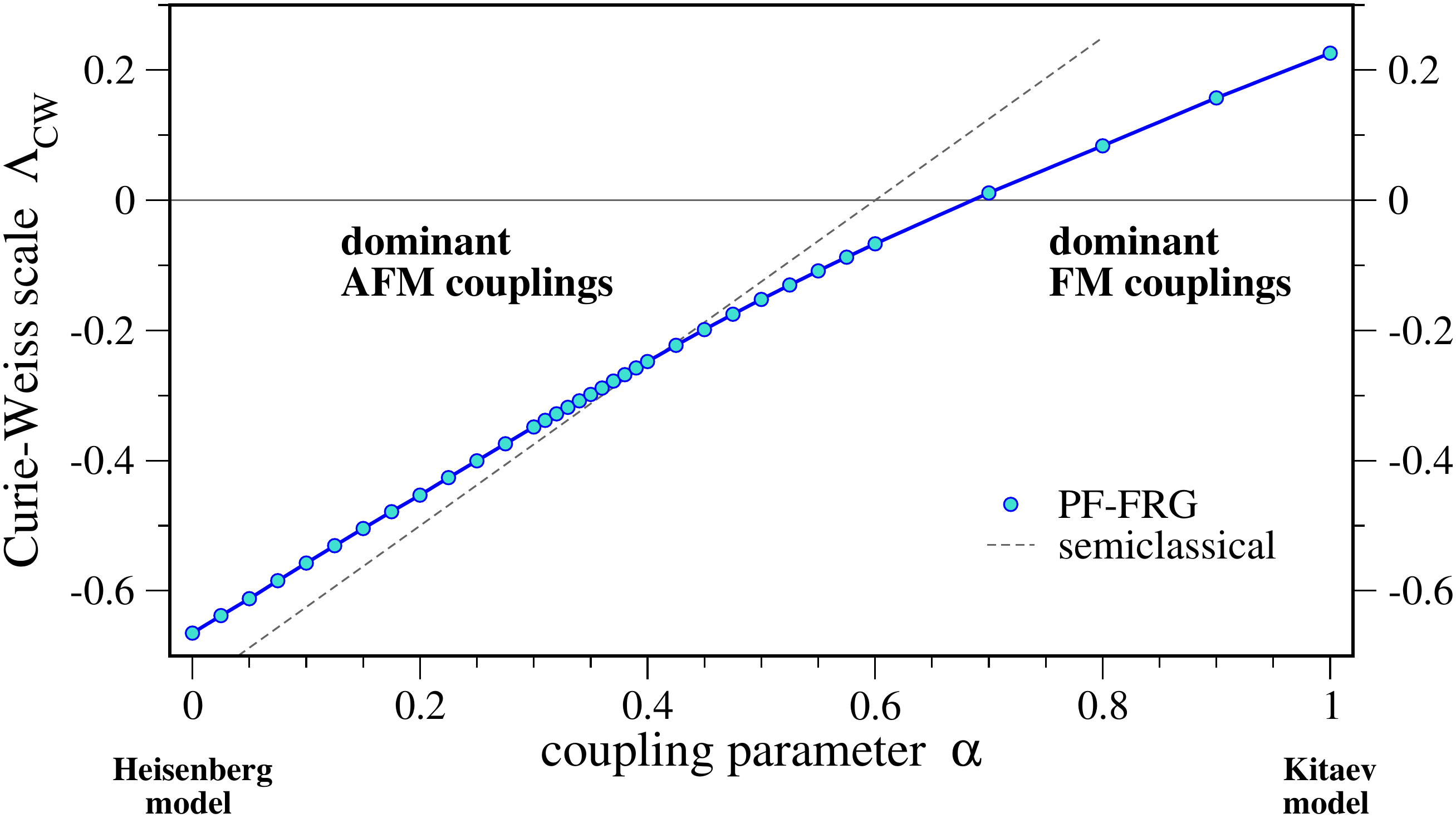}
   \caption{(color online)
              The Curie-Weiss scale \CW\ obtained from fitting the inverse 
              susceptibilities in Fig.~\ref{Fig:CW} to a Curie-Weiss law for
              varying coupling parameter $\alpha$.
              Around $\alpha \approx 0.67$ the Curie-Weiss scale
              switches sign indicating a transition of the dominant exchange
              from antiferromagnetic ($\Lambda_{\rm CW}<0$) to ferromagnetic ($\Lambda_{\rm CW}>0$). 
             }
   \label{Fig:CWT}
\end{figure}

We now turn to the finite-temperature properties of model \eqref{Eq:KitaevHeisenbergHamiltonian}.
To make a connection between the flow parameter $\Lambda$ and
the temperature $T$, we follow a line of thought first discussed by Honerkamp
and Salmhofer \cite{Honerkamp}: Both the flow parameter $\Lambda$ 
and the temperature $T$ act as infrared frequency cutoffs. While the former
is implemented as a sharp cutoff in the continuous frequency space, 
the latter allows a description in terms of discrete Matsubara frequencies,
where the smallest mesh point sets a lower bound of the energy resolution. 
Even though the precise cutoff procedures associated with $\Lambda$ and $T$ 
are hence quite different, we find that the identification of the two scales leads
to qualitatively correct results; quantitative uncertainties  possibly enter in our 
estimates of the ordering instability and its critical scale $\Lambda_c$. 

At high temperatures, we find that the homogeneous
susceptibility calculated from the RG flow for various $\Lambda$ nicely 
reproduces the expected Curie-Weiss behavior
\begin{equation}
\chi=C/(\Lambda-\Lambda_{\text{CW}}) \,,
\label{Eq:CurieWeiss}
\end{equation}
as shown in Fig.~\ref{Fig:CW}, which allows to extract rather precise numerical 
estimates for the Curie-Weiss scale $\Lambda_{\text{CW}}$, with the latter
being plotted in Fig.~\ref{Fig:CWT}.
Notably,  we observe that the Curie-Weiss scale changes sign
-- indicating a transition of the dominant exchange from antiferromagnetic to ferromagnetic -- 
around $\alpha \approx 0.68$, which is still deep in the magnetically ordered regime. 
Such a change of the dominant exchange is already suggestive from a semiclassical analysis 
of~\eqref{Eq:KitaevHeisenbergHamiltonian}, which gives $T_{\rm CW} = -3/4 + 5\alpha/4$
and thus indicates a sign change of the Curie-Weiss temperature around $\alpha=0.6$.
This further supports that the cutoff $\Lambda$ indeed retains the features of a temperature
parameter and justifies our assumption that $\Lambda$ can be used to
deduce finite-temperature properties of model \eqref{Eq:KitaevHeisenbergHamiltonian}. 

\begin{figure}[t]
    \includegraphics[width=\columnwidth]{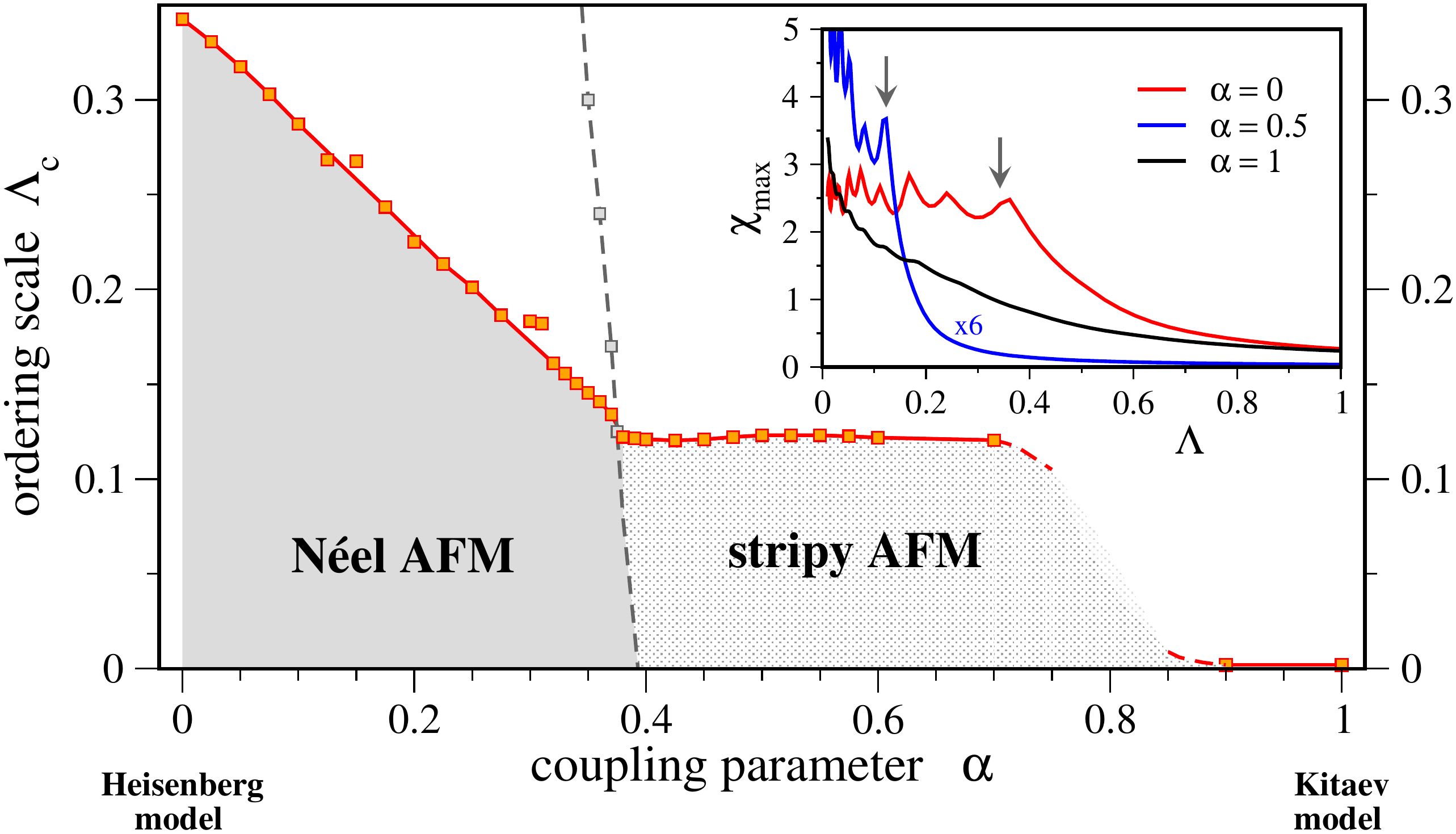}
   \caption{(color online)
                 Ordering scale $\Lambda_c$ obtained from the FRG calculations for various coupling parameters.
                 The dashed line indicates the crossover from dominant AFM to dominant s-AFM fluctuations as well as
                 an extrapolation below the ordering transition down to $T=0$. 
                 A regime of enhanced numerical uncertainties is encountered near $\alpha\approx0.8$.
                 The inset shows the RG flow of the magnetic susceptibility versus frequency cutoff $\Lambda$.
                 The arrows indicate the estimated ordering temperatures $\Lambda_c$ where the RG flow breaks down. }
   \label{Fig:OrderingTemperature}
\end{figure}

\begin{figure}[t]
    \includegraphics[width=\columnwidth]{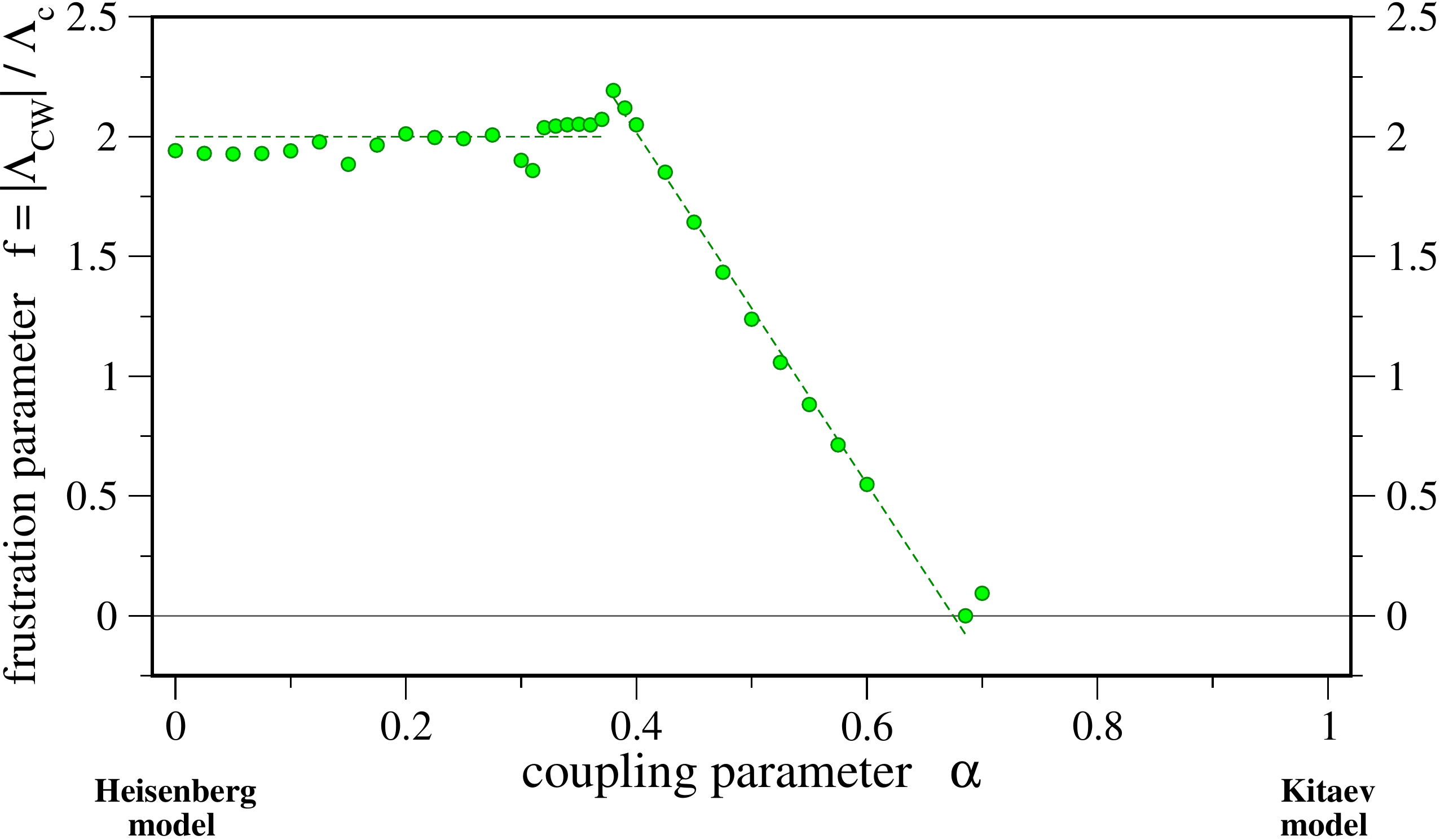}
   \caption{(color online)
                 The frustration parameter, i.e. the ratio of Curie-Weiss temperature \CW\ and ordering temperature $\Lambda_c$.}
   \label{Fig:FrustrationParameter}
\end{figure}

We can now return to the question whether a substantial frustration builds up as one 
interpolates between the spin-dominated (unfrustrated) Heisenberg regime to the orbital-dominated
(strongly frustrated) Kitaev regime. 
A commonly used measure for frustration is the ratio between Curie-Weiss and ordering
scale, the so-called frustration parameter
\begin{equation}
   f = |\Theta_{\text{CW}}|/T_c \approx |\Lambda_{\text{CW}}|/\Lambda_c \,,
   \label{Eq:Frustration}
\end{equation}
with a small value $f \lesssim 5$ indicating the absence of frustration, and systems with $f \gtrsim 10$
being commonly referred to as highly frustrated \cite{Ramirez}.
We estimate the ordering scale $\Lambda_c$ from the breakdown of the RG flow, as shown
in the inset of Fig.~\ref{Fig:OrderingTemperature}, with the result being plotted in the main panel 
of Fig.~\ref{Fig:OrderingTemperature} for the full parameter range $\alpha$ except for a region 
around the transition between the s-AFM phase and the SL $(\alpha \approx 0.8)$, 
where our approach does not allow to reliably calculate the transition temperature.
As shown in Fig.~\ref{Fig:FrustrationParameter} we observe a rather constant
plateau $f \approx 2$ for the frustration parameter in the AFM regime before it
decreases linearly starting around $\alpha \approx 0.4$ with the onset of the s-AFM phase.
At $\alpha=0.5$ in the s-AFM phase,
model~\eqref{Eq:KitaevHeisenbergHamiltonian} can be mapped to
a fluctuation-free classical system \cite{chaloupka-10prl027204}. This is consistent with our result of $f \approx 1$ 
at $\alpha=0.5$, as a frustration parameter close to unity signals the absence of 
fluctuation-induced frustration. 
For larger $\alpha$ the frustration parameter goes through zero as the Curie-Weiss scale
changes sign and beyond a regime of numerical uncertainty rapidly diverges as expected
for the spin-liquid regime, see Fig.~\ref{Fig:FrustrationParameter}.

\paragraph{\it Connection to experiments.--}

We close our discussion of the finite-temperature properties of model \eqref{Eq:KitaevHeisenbergHamiltonian}
by comparing our findings to recent thermodynamic measurements~\cite{singh-10prb064412} on the Iridate \Na.
The reported Curie-Weiss temperature of $\Theta_{\text{CW}} \approx -116$~K indicates a dominant AFM exchange
and the considerable suppression of magnetic ordering down to $T_{\text{N}} \approx 15$~K corresponds to a 
frustration of  $f \approx 8$.
While our finite-temperature analysis of the Heisenberg-Kitaev model \eqref{Eq:KitaevHeisenbergHamiltonian}
indicates an AFM Curie-Weiss temperature for a wide range of couplings $0 \leq \alpha \leq 0.68$, 
the ground states in this regime are relatively simple, magnetically ordered states that do not give rise to
a significant suppression of the ordering temperature with the frustration parameter $f$ never exceeding 
$f \approx 2$ in our calculations for this regime, see Fig.~\ref{Fig:FrustrationParameter}. 
On the other hand, we find a strong suppression of the ordering temperature in the spin-liquid phase for $\alpha > 0.8$
and its proximity, but the dominant couplings in this regime are ferromagnetic $(\Theta_{\text{CW}}>0)$.
To reconcile the combination of an AFM Curie-Weiss temperature and a simultaneous suppression of the ordering
temperature, one might thus want to look beyond the Heisenberg-Kitaev model \eqref{Eq:KitaevHeisenbergHamiltonian}.
In particular, one might want to consider various mechanisms that could suppress the ordering temperature in 
the magnetically ordered regime, such as a next-nearest neighbor exchange introducing geometric frustration as
suggested in Ref.~\onlinecite{jin-cm0907} or the role of disorder \cite{Trousselet}, especially in the form of non-magnetic 
impurities arising from the experimentally observed \cite{singh-10prb064412} site mixing between Ir and Na atoms. 
While the current analysis might suggest that \Na\  is not in close proximity to the spin liquid phase
of the Kitaev limit and its cousin -- a topological spin liquid in the presence of a magnetic field \cite{jiang-cm1101}, 
one might still speculate how one could drive the system closer to that regime.
One promising path to experimentally increase the anisotropic couplings might be to exert pressure 
along the $ab$-plane to counteract the $c$-axis lattice distortion in the material, which quenches the 
SOC. A similar relief of the lattice distortions might also be expected when replacing the Na ions by 
smaller Li ions \cite{JackeliPrivate} and consider \Li\ as a candidate material for more exotic ground states.

\begin{acknowledgements}
RT thanks G.~Jackeli for a particularly insightful discussion. We further
acknowledge discussions with H.-C. Jiang, R.~Moessner, and P.~W\"olfle. 
RT thanks all participants of the `Korrelationstage 2011' in Dresden for discussions.
JR thanks Microsoft Station Q for hospitality. JR is supported
by DFG-FOR 960, RT by the Humboldt Foundation.
\end{acknowledgements}


\end{document}